\documentclass[aps,nofootinbib,superscriptaddress,twocolumn]{revtex4-1}
\usepackage{graphicx}
\usepackage{amsmath,amssymb}
\usepackage{hyperref}
\usepackage[usenames]{color}

\hypersetup{
    colorlinks=true,
    linkcolor=red,
    citecolor=blue,
}

\def\be{\begin{equation}}
\def\ee{\end{equation}}
\def\ba{\begin{eqnarray}}
\def\ea{\end{eqnarray}}

\def\f{\frac}
\def\l{\left}
\def\r{\right}
\def\hub{{\cal H}}

\newcommand\etal{{et al}}
\newcommand\eg{{e.g.}}
\newcommand\ie{{i.e.~}}
\DeclareMathOperator{\adj}{adj}

\begin{document}

\title{A practical approach to cosmological perturbations in modified gravity}

\date{\today}

\author{Alessandra Silvestri}
\email{asilvest@sissa.it}
\affiliation{SISSA - International School for Advanced Studies, Via Bonomea 265, 34136, Trieste, Italy}
\author{Levon Pogosian}
\email{levon@sfu.ca}
\affiliation{Department of Physics, Simon Fraser University, Burnaby, BC, V5A 1S6, Canada}
\affiliation{Centre for Theoretical Cosmology, DAMTP, University of Cambridge, CB3 0WA, UK}
\author{Roman V. Buniy}
\email{roman.buniy@gmail.com}
\affiliation{Schmid College of Science, Chapman University, Orange, CA 92866, USA}

\begin{abstract}
The next generation of large scale surveys will not only measure cosmological parameters within the framework of General Relativity, but will also allow for precision tests of the framework itself. At the order of linear perturbations, departures from the growth in the LCDM model can be quantified in terms of two functions of time and Fourier number $k$. We argue that in local theories of gravity, in the quasi-static approximation, these functions must be ratios of polynomials in $k$, with the numerator of one function being equal to the denominator of the other. Moreover, the polynomials are even and of second degree in practically all viable models considered today. This means that, without significant loss of generality, one can use data to constrain only five functions of a single variable, instead of two functions of two variables. Furthermore, since the five functions are expected to be slowly varying, one can fit them to data in a non-parametric way with the aid of an explicit smoothness prior. We discuss practical application of this parametrization to forecasts and fits.
\end{abstract}

\maketitle

\section{Introduction}

Upcoming redshift and weak lensing surveys, such as Dark Energy Survey (DES)~\cite{des}, Euclid~\cite{euclid} and Large Synoptic Survey Telescope (LSST)~\cite{lsst,Ivezic:2008fe}, combined with the cosmic microwave background measurements from Planck~\cite{planck} and other cosmological probes, will accurately trace the growth of cosmic structures through multiple epochs. They will offer the opportunity to test General Relativity (GR) by examining the relations between the distribution of matter, the gravitational potential and the lensing potential on cosmological scales. Such tests may yield clues to the physics causing cosmic acceleration or, at the very least, extend the range of scales over which Einstein's gravity has been validated by experiment. 

To test GR, one can either constrain particular alternative gravity models, such as the Dvali-Gabadadze-Porrati (DGP) braneworld model~\cite{Dvali:2000hr} or $f(R)$~\cite{Capozziello:2003tk,Carroll:2003wy}, or work within more general parametrized frameworks that cover many theories at once and minimize the risk of missing potential hints of modified gravity in the data. Over the past several years significant effort went into developing such frameworks and understanding requirements for their consistency~\cite{Linder:2007hg,Zhang:2007nk,Hu:2007pj,Bertschinger:2008zb,Daniel:2008et,Skordis:2008vt,Song:2010rm,Bean:2010zq,Daniel:2010ky,Pogosian:2010tj,Baker:2011jy,Battye:2012eu,Sawicki:2012re,Baker:2012zs,Gubitosi:2012hu,Amendola:2012ky,Bloomfield:2012ff}. Often, departures from the standard cosmological model (LCDM) are quantified in terms of arbitrary functions of time and, sometimes, scale. These functions cannot be fit to data without further assumptions about their form.

In this paper we motivate a parametrization that contains five unknown functions of time only and is general enough to cover most viable models of modified gravity and dark energy proposed so far. Importantly, these functions are expected to be slowly varying, hence the effective number of degrees of freedom that are fit to data can be small. One can avoid assuming a parametric form for the five functions and use instead a smoothness prior similarly to how it was applied to reconstruction of the dark energy equation of state $w$ in~\cite{Crittenden:2005wj,Crittenden:2011aa,Zhao:2012aw}.

Observables describing large scale structure are calculated using cosmological perturbation theory in Fourier space. The relevant variables are the two scalar metric degrees of freedom, \eg, $\Phi$ and $\Psi$ in the Newtonian gauge, along with the matter density contrast $\delta$ and the matter velocity perturbation $v$. One needs four equations to solve for the evolution of these four variables, assuming that baryons and dark matter obey the same equations at late times. Two equations are provided by the covariant conservation of matter energy-momentum. The other two equations are supposed to be provided by a theory of gravity which prescribes how the metric responds to the matter stress-energy. 
Formally, one can always complete the system of equations by introducing two functions $\mu(a,k)$ and $\gamma(a,k)$, defined via\footnote{These definitions assume that anisotropic stress of matter is negligible at the epochs of interest, although it can be included, if necessary, as it was done in~\cite{Bean:2010zq,Hojjati:2011ix}.}
\be
k^2 \Psi =- 4 \pi\mu G a^2 \rho \Delta \ , \quad \Phi = \gamma \Psi \ ,
\label{eq:mugamma}
\ee
where $a$ is the scale factor and $\Delta=\delta+3aHv/k$. They are defined in a way that recovers the Poisson and the anisotropy equations of LCDM when $\mu=\gamma=1$. There are other choices in the literature for the pair of functions relating ($\Phi$, $\Psi$) to ($\delta$, $v$) that are equivalent to $\mu$ and $\gamma$. A common alternative choice is to use $\Sigma$, defined as
\be\label{Poisson_Weyl_Modified}
k^2\l(\Phi+\Psi\r)=- 8\pi G a^2 \Sigma(a,k) \rho\Delta\ ,
\ee
in combination with $\mu$. As shown in~\cite{Zhao:2008bn,Pogosian:2010tj,Bean:2010zq,Hojjati:2011ix}, once the two functions are given, one has a consistent set of equations that can be incorporated~\cite{Zhao:2008bn,Hojjati:2011ix,mgcamb,Dossett:2011tn} into standard Boltzmann codes, such as CAMB~\cite{Lewis:1999bs}, to calculate the observables. In principle, everything that observations can tell us about cosmic structure on linear scales can be stored as a measurement of $\mu$ and $\gamma$ and, if necessary, projected onto constraints on specific models.

But what form should one adopt for these functions to fit them to data? In \cite{Zhao:2009fn,Hojjati:2011xd,Hall:2012wd}, a principal component analysis (PCA) was performed to forecast the best constrained eigenmodes of $\mu(a,k)$ and $\gamma(a,k)$ for different future surveys, finding that they will measure amplitudes of tens (if not hundreds) of them with good accuracy. This is encouraging, but it is not clear how many of these constrainable eigenmodes are physically interesting. PCA alone does not really answer the question of what parameters one should be fitting to data.

Another concern, which is the main motivation for this work, is that an arbitrary relation between two quantities in Fourier space, such as those in Eq.~(\ref{eq:mugamma}), does not, in general, imply a local relation between them or their derivatives in real space.
Clearly, the $k$-dependence of $\mu(a,k)$ and $\gamma(a,k)$ cannot be completely arbitrary if equations of motion are obtained from variational principle.

In this work, we investigate the physically acceptable forms of $\mu(a,k)$ and $\gamma(a,k)$ based on considerations of locality and general covariance. We show that under rather general conditions, and under the quasi-static approximation (QSA), they should always have a form of ratios of polynomials in $k$. Furthermore, the numerator of $\mu$ is set by the denominator of $\gamma$. The coefficients inside the polynomials are functions of the background quantities and can be expected to be slowly varying functions. Technically, the number of these time-dependent coefficients is infinite if one allows for a completely arbitrary modification of GR.  However, in models with purely scalar extra degrees of freedom, the polynomials are even in $k$ and, furthermore, in many viable models considered so far in the literature, the polynomials are even and of second degree, hence the number of time-dependent coefficients is reduced to five. While this parametrization is motivated by the QSA, it allows for departures form LCDM on near-horizon scales. 

Baker \etal. \cite{Baker:2012zs} have recently investigated the form of exact equations of motion for a large variety of modified gravity models. They also noted that, under the QSA, equations reduce to algebraic relations with even powers of $k$ and proposed constraining the time dependence of the background dependent coefficients. Instead, we consider coefficients appearing in $\mu$ and $\gamma$, which reduces the number of free functions in most interesting cases to five and makes it easy to use them in existing modified Boltzmann codes, such as MGCAMB \cite{Zhao:2008bn,Hojjati:2011ix,mgcamb}. Amendola \etal. \cite{Amendola:2012ky} recently adopted an equivalent five-function parametrization to investigate limits of observability of modified gravity on linear scales. Their choice was motivated by results of De Felice \etal. \cite{DeFelice:2011hq}, who calculated $\mu$ and $\gamma$ in the QSA for the Horndeski \cite{Horndeski:1974wa} class of most general second order scalar-tensor theories. We arrive at the same form as a particular case of a more general and much simpler derivation.

Working with five arbitrary functions of time may seem like a daunting task, but it is much easier than constraining two functions of scale and time. Furthermore, the functions are known to be slowly varying, which can be used as a strong theoretical prior. We outline the practical application to forecasts and fits in Sec.~\ref{sec:app}. Our parametrization is useful if one wants to look for departures from LCDM without assuming a particular model.  Clearly, the number of functions can be smaller if one restricts the range of possibilities. For instance, to describe linear perturbations in Brans-Dicke models it is sufficient to provide only two functions of the background~\cite{Brax:2011aw,Brax:2012gr}.

The remainder of the paper is organized as follows. In Sec.~\ref{sec:general}, we show that, under the QSA, $\mu$ and $\gamma$ are ratios of polynomials in $k$, and the numerator of $\mu$ is given by the denominator of $\gamma$. In \ref{sec:viable} we point out that for a very broad class of viable modified gravity models, the polynomials are even and second order in $k$ and, therefore, one needs to specify only five functions of time. In Sec.~\ref{sec:QS}, we examine the assumptions made by the QSA and discuss the extent of their applicability. In Sec.~\ref{sec:app}, we outline the procedure for reconstructing $\mu$ and $\gamma$ from data using a smoothness prior applied to the five functions. We conclude with a discussion in Sec.~\ref{sec:summary}.

\section{$\mu$ and $\gamma$ in modified gravity models}
\label{sec:mugamma}

\subsection{The most general case}
\label{sec:general}

Consider a broad class of theories in $(3+1)$ dimensions with the action defined in terms of a Lagrangian density that contains an arbitrary function of geometric invariants $R$, $R_{\alpha\beta} R^{\alpha\beta}$, $R_{\alpha\beta\gamma\delta}R^{\alpha\beta\gamma\delta}$, $\Delta R$, $R^{\alpha\beta}\nabla_\alpha\nabla_\beta R,\dotsc$ as well as any number of scalar degrees of freedom $\phi_i$, $i=1,\dotsc,N$ (including the longitudinal components of vector or tensor degrees of freedom), which can be non-minimally coupled to the metric and each other. This embraces dark energy models as well as modified gravity theories, including effective $(3+1)$ dimensional descriptions of higher dimensional theories. For the moment, let us not worry about the existence of ghosts or other unphysical properties that such theories may have. At this point, we only require invariance of the action under general coordinate transformations. Let us also make an important assumption that there exists a frame in which all particles are minimally coupled to the metric, so that the matter stress-energy is covariantly conserved. For simplicity, we neglect radiation and the differences between CDM and baryons.

Let us now consider the form of equations for linear scalar perturbations in the Newtonian gauge, where the relevant degrees of freedom of the metric sector are the potentials $\Phi$ and $\Psi$ defined as
\be
ds^2=-(1+2\Psi)a^2d\tau^2+(1-2\Phi)a^2d{\bf x}^2 \ .
\ee
Varying the action with respect to the metric tensor gives four Einstein equations. The time-time and the time-space components can be combined to form the Poisson equation, which, to linear order in the perturbations in Fourier space, will have the following general form:
\be
{\hat A} \Psi + {\hat B} \Phi + {\hat C}^i \delta \phi_i = -4 \pi G a^2 \rho \Delta \ ,
\label{eq:poisson}
\ee
where we assume summation over repeated indices and where ${\hat A}$, ${\hat B}$ and ${\hat C}^i$ are linear operators that contain functions of the background, time derivatives and/or powers of $k$.
For instance, for any local theory of gravity, one generally has
\be
{\hat A} = \sum_{n,m}a_{nm}k^{n}\partial^m_0 \ ,
\label{eq:ahat}
\ee
where $\partial^m_0$ denotes the time ($\tau$) derivative of $m$th order,  the highest values of $n$ and $m$ are determined by the order of metric derivatives contained in the action, and the coefficients $a_{nm}$ are functions of time. Note, that in models with only scalar extra degrees of freedom, ${\hat A}$ will contain only {\it even} powers of $k$. This is because odd powers can only come from a contraction of spatial derivatives of perturbations with spatial indices of the background dependent coefficients that should vanish for isotropic backgrounds such as FRW. 
Operators ${\hat B}$ and ${\hat C}^i$ have the same form with corresponding coefficients $b_{nm}$ and $c_{nm}^i$.

Similarly, the traceless space-space Einstein equation can generally be written as
\be
{\hat D} \Psi + {\hat E} \Phi + {\hat F}^i \delta \phi_i = 0 \ ,
\label{eq:ij}
\ee
where the zero on the right hand side is due to vanishing of the matter anisotropic stress, and the operators ${\hat D}$, ${\hat E}$ and ${\hat F}^i$ have the same form as ${\hat A}$ in Eq.~(\ref{eq:ahat}). In addition, varying the action with respect to each scalar field $\phi_i$, and linearizing in the perturbations, will provide equations for the corresponding perturbation $\delta\phi_i$, that can generally be written as
\be
{\hat H}_i \Psi + {\hat K}_i \Phi + {\hat L}_i^{\ j} \delta \phi_j = 0 \ ,
\label{eq:phi}
\ee
where the operators ${\hat H}_i$, ${\hat K}_i$ and ${\hat L}_i^{\ j}$ also have the form given by Eq.~(\ref{eq:ahat}), with correspondingly renamed coefficients.

Our aim is to find the form of the functions $\mu(a,k)$ and $\gamma(a,k)$ defined in Eq.~(\ref{eq:mugamma}). Because of the time-derivatives in Eqs.~(\ref{eq:poisson}), (\ref{eq:ij}) and (\ref{eq:phi}), it is impossible to write $\mu(a,k)$ and $\gamma(a,k)$ in a closed form without solving for the evolution of the perturbations first. To make progress, let us take the quasi static approximation (QSA) in which we neglect all time derivatives of $\Phi$, $\Psi$ and $\delta \phi_i$, and delegate justifying this approximation to Section \ref{sec:QS}.

In the QSA, the operators ${\hat A}$, $\hat{B}$, $\hat{C}^i$, $\hat{D}$, $\hat{E}$, $\hat{F}^i$, $\hat{H}_i$, $\hat{K}_i$, $\hat{L}_i^{\ j}$ become functions, specifically polynomials in $k$; we indicate them with the same letters, removing the hats, \eg,  we have
\begin{align}
  A = \sum_{n}a_{n0}k^{n} \ 
  \label{}
\end{align}
and similarly for the other functions.
As a result, Eqs~(\ref{eq:poisson}),~(\ref{eq:ij}) and~(\ref{eq:phi}) reduce to a system of linear algebraic equations which we can use to extract $\mu$ and $\gamma$.

Defining $R_i$ via
\be
\delta \phi_i = R_i \Psi
\ee
and substituting it, along with $\Phi=\gamma \Psi$, into Eqs.~(\ref{eq:ij}) and (\ref{eq:phi}), we find
\ba
\label{eq:gamma-R1}
D + E \gamma  + F^i R_i &=& 0 \ , \\
H_i + K_i \gamma + L_i^{\ j} R_j &=& 0 \ .
\label{eq:gamma-R2}
\ea
It is convenient to write the solution of these equations in the matrix form,
\begin{align}
  \begin{bmatrix}
    \gamma \\
    R
  \end{bmatrix}
  = -
  \begin{bmatrix}
    E & F \\
    K & L
  \end{bmatrix}
  ^{-1}
  \begin{bmatrix}
    D \\
    H
  \end{bmatrix}
  ,
  \label{}
\end{align}
where we introduced a row vector $F$, a column vector $K$ and a square matrix $L$. We can express the inverse of any matrix $M$ as the ratio of its classical adjoint to its determinant, $M^{-1}=\adj{M}/\det{M}$. After some algebra we obtain
\begin{alignat}{2}
  &\det
  \begin{bmatrix}
    E & F \\
    K & L
  \end{bmatrix}
  &&=(E-FL^{-1}K)\det{L}\nonumber\\&&&=E\det{(L-KE^{-1}F)}\ ,
  \label{det}\\
  &\adj
  \begin{bmatrix}
    E & F \\
    K & L
  \end{bmatrix}
  &&=
  \begin{bmatrix}
    \det{L} & -F\adj{L} \\
    -(\adj{L})K & E\adj{(L-KE^{-1}F)}
  \end{bmatrix}
  \ .
  \label{adj}
\end{alignat}
Since $D$, $E$, $F$, $H$, $K$, $L$ are polynomials, the quantities in \eqref{det} and \eqref{adj} are polynomials as well, and, consequently, $\gamma$ and $R$ are fractions of polynomials. Furthermore, in their irreducible form, the denominators of $\gamma$ and $R$ are the same,
\be\label{gamma_and_R_pol}
\gamma=\frac{N_\gamma}{Q} \ ,\,\,\,\,
R=\frac{N_R}{Q} 
\ee
where 
\begin{align}
  N_\gamma&=-D\det{L}+F(\adj{L})H\ ,\\
  N_R&=D(\adj{L})K-E\,{\rm adj}\,(L-KE^{-1}F)H\ ,\\
  Q&=E \det{(L-K E^{-1}F)} \ .
  \label{}
\end{align}
Also, since the Poisson equation in the QSA has the form
\be
A + B \gamma + C^i  R_i = {-4 \pi G a^2 \rho \Delta \over \Psi} = \frac{k^2}{\mu}\ ,
\ee
it follows that $\mu$ can be written as an irreducible fraction of polynomials with its numerator uniquely determined by the denominator of $\gamma$ and $R$, \ie
\be\label{mu_pol}
\mu=\frac{k^2 Q}{AQ+B\,N_\gamma+CN_R}\ .
\ee

As mentioned earlier, in models with pure scalar degrees of freedom, the polynomial functions will contain only even powers of $k$. Furthermore, the $k^0$ terms in the denominator of $\mu$ are negligible in the QSA because the magnitude of the corresponding coefficients in the Poisson equation is dependent either on $H$, time-derivatives of $\phi_i$ or the first derivative of the scalar field potential(s), all of which are small in the QSA. This means that the $k^2$ factors in the numerator and denominator of $\mu$ will cancel.

Thus, starting from a general covariant action that contains an arbitrary function of geometric invariants and any number of scalar degrees of freedom (DOF), we derived in a model-independent way that, under the QSA, the functions $\gamma$ and $\mu$ are ratios of polynomials in $k$, with the numerator of $\mu$ equal to the denominator of $\gamma$. 

\subsection{The subset of viable models}
\label{sec:viable}
A parametrization that anticipates a completely arbitrary modification of gravity is impractical, as one cannot fit an infinite number of unknown functions to data. Furthermore, there are good theoretical reasons not to allow for arbitrarily high order derivatives or tensorial modes in the equations of motion because of the appearance of ghost degrees of freedom. Let us therefore consider a more representative class of viable modified gravities described by a Lagrangian that contains only one scalar DOF obeying second order equations of motion. In this case, 
Eqs~(\ref{eq:poisson}), (\ref{eq:ij}) and (\ref{eq:phi}) in the QSA reduce to:
\ba
&&A k^2 \Psi + B k^2 \Phi + C k^2 \delta \phi = -4 \pi G a^2 \rho \Delta\ , \\
&&D \Psi + E \Phi + F \delta \phi = 0\ , \\
&&H k^2 \Psi + K k^2 \Phi + (L_0 + L_1 k^2) \delta \phi = 0 \ ,
\ea
where we have made the $k$ dependence explicit, so that $A$, $B$, $\dotsc$, $L_1$ are time dependent coefficients, and we include the $L_0$ coefficient which represents the mass squared of the scalar field.

Following the same steps as in Sec.~\ref{sec:general} we find in this case that $\mu$ and $\gamma$ are ratios of {\em even}  polynomials of {\em second degree} and, as in the general case, the denominator of $\mu$ is the same as the numerator of $\gamma$, i.e. 

\ba\label{gamma_pol_2nd}
\gamma&=&\frac{-DL_0+\left(FH-DL_1\right)k^2}{EL_0+\left(EL_1-KF\right)k^2}\ ,\\
\label{mu_pol_2nd}
\mu&=&\left[EL_0+(EL_1-KF)k^2\right]\Bigl\{(AE-BD)L_0\nonumber\\ 
&& \quad\quad +\bigl[F(BH
-AK)+C(KD-EH)\nonumber\\ && \quad\quad +(AE-BD)L_1\bigr]k^2\Bigr\}^{-1}.
\ea
The above expressions have the same forms as analogous expressions derived in~\cite{DeFelice:2011hq} for general Horndeski theories \cite{Horndeski:1974wa}. Indeed, although we arrived at~(\ref{gamma_pol_2nd}) and (\ref{mu_pol_2nd}) from general arguments, the subset of viable models to which we are restricting coincides with the models included in the Horndeski class, which contains most of the viable theories of dark energy and modified gravity. The class of theories with a single scalar DOF with a second order equation of motion includes models with actions that contain a function $f(R,G)$ of the Ricci scalar and the Gauss-Bonnet term, provided the determinant of the Hessian is zero, i.e. $f_{RR}f_{\mathcal{G}\mathcal{G}}-f^2_{R\mathcal{G}}=0$. Restricting to the Lovelock invariants~\cite{Lovelock:1971yv} $R$ and $\mathcal{G}$ guarantees that no spurious spin-2 ghosts are introduced~\cite{DeFelice:2010aj}, while having a null determinant of the Hessian further ensures that superluminal modes for scalar perturbations are avoided~\cite{DeFelice:2010hb}. Finally, Horndeski theories include also dark energy models such as quintessence and k-essence, as well as the covariant Galileon and the 4 dimensional effective DGP model in the decoupling limit~\cite{Nicolis:2008in}. 

Without loss of generality, we can rewrite~(\ref{gamma_pol_2nd}) and (\ref{mu_pol_2nd}) in a more compact way by introducing $5$ functions of the background $\{p_i(a)\}$:
\ba
\label{eq:gamma}
\gamma &=&  {p_1(a)+p_2(a) k^2 \over 1+p_3(a) k^2}\ , \\
\mu &=& {1 + p_3(a) k^2 \over p_4(a) + p_5(a) k^2} \ .
\label{eq:mu}
\ea
Thus, in the QSA, one can express the perturbed equations of motion of a very large class of viable modified gravity models in terms of only 5 functions of time.

Note that even though this ansatz was derived using the QSA, it allows for near- and super-horizon modifications of gravity: $\gamma(a,k\rightarrow 0)=p_1(a) \ne 1$. Also note that, while $\mu$ can also deviate from unity on super-horizon scales ($\mu \rightarrow p_4^{-1}(a)$), this should not affect any of the observables and the  super-horizon perturbations will evolve consistently with the background expansion~\cite{Pogosian:2010tj}.

Analogous compact forms (\ref{eq:gamma}) and (\ref{eq:mu}) were used in~\cite{Amendola:2012ky}, based on results in \cite{DeFelice:2011hq}, to discuss prospects of constraining Horndeski theories~\cite{Horndeski:1974wa}. We arrive at the same forms starting from simpler and more general arguments that do not require considering the details of the Horndeski action. Finally, the form of $\gamma$ and $\mu$ in~(\ref{eq:gamma}) and (\ref{eq:mu}) resembles the parametrization introduced by Bertschinger and Zukin (BZ) in~\cite{Bertschinger:2008zb}, however, there are some important differences. In the  BZ parametrization, $\mu$ and $\gamma$ tend to $1$ for $k\rightarrow 0$, effectively reducing the theory to GR in that limit. Furthermore, BZ does not set the numerator of $\mu$ equal to the denominator of $\gamma$, which we find instead to be a general feature. Finally, they fix the time dependence of the coefficients in the $k^2$ expansion to a power law, while we leave it general. From a theoretical point of view, not all time dependencies can be described as power laws; however, the differences might be undetectable depending, in part, on the range of redshifts probed by the experiment.

\section{The quasi-static approximation}
\label{sec:QS}
As mentioned earlier, closed form expressions for $\mu(a,k)$ and $\gamma(a,k)$ may not exist in a general gravity theory unless one adopts the QSA, in which the relations between the metric potentials and the matter perturbation become algebraic. Without the QSA, one needs to solve the differential equations in order to determine $\mu$ and $\gamma$, making them dependent on the initial conditions. 

There are {\it two} different assumptions involved in what is commonly known as the QSA: (1) the relative smallness of the time derivatives of metric perturbations compared to their space derivatives, and (2) the sub-horizon approximation, $k/aH \gg 1$. In LCDM, the second assumption automatically implies the first --- the perturbed quantities evolve on time scales comparable to the expansion rate, thus their time derivatives become comparable to space derivatives only for perturbations on near-Hubble scales. In alternative gravity models with additional degrees of freedom, the two assumptions need not imply each other, so let us separately consider their effects on the applicability of our parametrization.

\subsection{Neglecting time-derivatives} 

In scalar-tensor models of modified gravity, there can be rapid oscillations of the metric potentials on top of their slow evolution which can make their time-derivatives large. A general solution typically includes a homogeneous oscillatory mode as well as a particular solution, that can also oscillate, induced by the coupling to matter. The initial amplitude of the former is a free parameter that typically needs to be fine-tuned to a small value in order to have a consistent cosmology at early times and avoid problems such as particle overproduction~\cite{Starobinsky:2007hu}. Motivated by this, we propose a theoretical prior in which the amplitude of the homogeneous mode is very small initially. Subsequently, it is controlled by functions of the background and can grow only slowly, never becoming large enough to affect observables. The amplitude of the oscillations in the particular solution is not a free parameter, but in general the terms that set the amplitude and the frequency of such oscillations are proportional to the strength of the coupling and the range of the extra scalar degree of freedom. Because of the generally tight constraints on fifth forces, one typically finds that the oscillations are undetectable. 

To illustrate the point, let us consider models in which the Einstein-Hilbert part of the action is given by $R+f(R)$, and the Poisson equation has the form~\cite{Pogosian:2007sw}
\ba
\label{Poisson_Psi_f(R)}
k^2\Psi&+&k^2\frac{\delta f_R}{2F}+\frac{3}{2}\l[\l(\dot{\hub}-\hub^2\r)\frac{\delta f_R}{F}+\l(\dot{\Phi}+\hub\Psi\r)\frac{\dot{F}}{F}\r]\nonumber\\
&&=-4\pi Ga^2\f{\rho}{F}\Delta \,,
\ea
where $f_R\equiv df/dR \equiv F-1$ is the ``scalaron'' degree of freedom, $\hub=a^{-1}da/d\tau$, and the non-quasi-static terms are collected inside the square brackets. On sub-horizon scales (when $k \ll aH$), the first term inside the square brackets is negligible compared to the $k^2 \delta f_R$ term. But on large scales it is {\it still} smaller than other terms because $\delta f_R=f_{RR}\delta R$, and $f_{RR}$ must be small. The latter is required for the Chameleon mechanism~\cite{Khoury:2003aq} to screen the fifth force inside our solar system --- the values of $f_R$ and $f_{RR}$ must be small~\cite{Hu:2007nk}. The second term inside the square brackets is large only if $\dot{\Phi}$ is large. However, the evolution of $\Phi$ in the heavy scalaron  (small $f_{RR}$) limit is practically the same as in GR, except for additional oscillations with a tiny amplitude set by $f_{RR}$~\cite{Hojjati:2012rf}. Even if the oscillations had a larger amplitude, they would be difficult to detect because of their high frequency set by $f^{-1}_{RR}$. Furthermore, there are no oscillations in the lensing potential $\Phi+\Psi$, hence there can be no signal in the Integrated Sachs-Wolfe effect that constrains the near- and super-horizon evolution of perturbations.

We are not aware of a theory in which oscillations in extra DOFs are observable for the range of parameters that has not already been ruled out. Thus, it is reasonable to {\it adopt a theoretical prior} that ignores rapid time-variations of gravitational degrees of freedom {\it until} we find an example of a viable theory in which they are observable. Finding such an example may warrant an appropriate extension of Eqs.~(\ref{eq:gamma}) and (\ref{eq:mu}).

\subsection{The sub-horizon approximation}

By ignoring the time derivatives in the modified equations we are neglecting not just the rapid oscillations in metric perturbations but also the slowly varying signatures of modified gravity. This, in the absence of additional information about the model, can only be justified in the $k/aH \gg 1$ limit. 

Before addressing the significance of near-horizon modifications of gravity, let us make an important point: the {\it implementation} of our parametrization in the equations of motion does not assume the QSA. In the LCDM limit, when $\mu=\gamma=1$, we recover the {\em exact} equations of GR, while the parametrization allows for departures from $\mu=\gamma=1$ on all scales. Also, we do not suggest that one should ignore the relativistic effects when calculating the observables~\cite{Yoo:2009au,Yoo:2010ni,Challinor:2011bk,Bonvin:2011bg,Bruni:2011ta,Jeong:2011as,Yoo:2013tc}. The implementation of near-horizon and other relativistic effects~\cite{Challinor:2011bk} in Boltzmann codes like CAMB is unaffected by the use of the $(\gamma,\mu)$ parametrization --- only the Einstein equations are modified, while Boltzmann equations and the expressions for the observable quantities remain the same as before.

The two relevant questions about the validity of taking the QSA limit in deriving the form of our parametrization are: (1) How observable departures from the LCDM prediction can be on near horizon scales in viable modified gravity models? and (2) To what extent our parametrization would bias such a potentially observable signature?

As far as we know, there is no example of a theoretically motivated model that is not ruled out and in which departures from LCDM on near-horizon scales have been shown to be detectable~\cite{Lombriser:2013aj}, although there is clearly more room for investigation. Part of the reason is that cosmic variance limits the statistical significance of any inference on large scales. Multi-tracer technique proposed in~\cite{Seljak:2008xr,Seljak:2009af,Yoo:2012se} can remove this limitation to some extent. However, in order for the modified gravity signal on large scales to be detectable, it has to be sufficiently pronounced while still keeping the model in agreement with other constraints. While one can design specific models providing such an example~\cite{Lombriser:2013aj}, they cannot be considered representative.

The second question --- the extent to which our parametrization would bias a potentially important signature on near-horizon scales --- can only be answered by considering particular solutions of specific models. But first one has to find a viable model in which such signatures are observable at all.

A conservative way to use the parametrization in Eqs.~(\ref{eq:gamma}) and (\ref{eq:mu}) would be to separately fit to a subset of data corresponding to clustering on sub-horizon scales. Then, if a departure from LCDM is seen, one would have a clear idea about which scales contribute the most to the signal, and whether it is appropriate to interpret it under the QSA.

\section{Practical application to forecasts, constraints and reconstructions}
\label{sec:app}

It is impossible to constrain a function without assuming something about its form. One possibility is to pick a particular functional form, such as a power law dependence, similar to how it is done in the BZ parametrization~\cite{Bertschinger:2008zb}. Since the five functions in Eqs.~(\ref{eq:gamma}) and (\ref{eq:mu}) are expected to be slowly varying, this need not be a bad approximation if the data only probes a limited range of redshifts. But it is unlikely that a single power law will capture the evolution over a wide range of epochs, which is what we can expect from surveys like Euclid, SKA or LSST in combination with CMB and other data. Let us instead explore a way to constrain $\mu$ and $\gamma$ in a non-parametric way that still takes into account their smoothness. 

Recent Refs. \cite{Crittenden:2011aa,Zhao:2012aw} proposed a transparent Bayesian framework for constraining the dark energy equation of state $w(a)$ based on adopting an explicit smoothness prior. The prior is defined via a correlation function that correlates values of $w$ at neighbouring points in $a$. This framework can be applied to any unknown function (or functions) expected to be smooth from theoretical considerations. Let us outline how it can be applied to $\{p_i(a)\}$ in Eqs.~(\ref{eq:gamma}) and (\ref{eq:mu}) for the purpose of forecasting future constraints on $\gamma$ and $\mu$, as well as for fitting to real data.

\subsection{Application to forecasting}

As a starting point, one can discretize the functions into finite numbers of bins. The binning can be implemented using a smooth function, such as a hyperbolic tangent, to avoid infinite derivatives at the edges, and the number of bins can always be taken to be sufficiently large to achieve convergence\footnote{One of the advantages of the smoothness prior approach of~\cite{Crittenden:2005wj,Crittenden:2011aa,Zhao:2012aw} is that it eliminates the dependence on binning.}. The binned values of $\{p_i(a)\}$ can be substituted into Eqs.~(\ref{eq:gamma}) and (\ref{eq:mu}) to find $\mu$ and $\gamma$ which are used as input in a modified Boltzmann code such as MGCAMB~\cite{mgcamb}. Along with providing other cosmological parameters, this is sufficient for calculating all types of cosmological observables.

In the simplest approach to forecasting, one assumes a Gaussian shape of the parameter likelihood surface with a  peak corresponding to a fiducial model, and proceeds to calculate the Fisher matrix from derivatives of observables with respect to model parameters. 
One then inverts it to obtain an estimate of the total covariance matrix. Because of the large number of highly correlated parameters, considering a constraint on any single bin is meaningless. One can instead use the principal component analysis (PCA)~\cite{Huterer:2002hy} to see which independent linear combinations of bins will be best constrained by a given experiment. This is accomplished by diagonalizing the corner of the covariance matrix corresponding to the bins of the five functions.

What can one do with the information obtained from the PCA forecast for $\{p_i(a)\}$? There will be a strong degeneracy between the five functions, which means one should 
look at independent linear combinations of bins of {\it all five} functions. 
The number of such well-constrained {\it combined} eigenmodes should give us a measure of how many physically relevant independent degrees of freedom one can measure about $\mu$ and $\gamma$.

In the absence of a theoretical prior, all eigenmodes of $\{p_i(a)\}$ carry some information. How should one decide which modes are informative and which are not? This is where one can use the knowledge about the slowly varying nature of the functions and introduce a smoothness prior~\cite{Crittenden:2005wj,Crittenden:2011aa,Zhao:2012aw}. The prior comes in a form of a non-diagonal matrix $C^{\rm prior}$, the inverse of which one adds to the inverse of the original covariance matrix, and which introduces additional correlations between bins of the five functions. One then considers an eigenmode to be informative if it is unaffected by the prior, \ie if it is the same before and after addition of the prior covariance.

To be more explicit, let us assume that each function $p_i(a)$ is binned into $N$ bins in the scale factor, $a_1,\dotsc,a_N$, and label the bins so that they form a single vector ${\tilde p}_\alpha$ with $\alpha=1,\dotsc,5N$. For example, one can define
\begin{align}
 {\tilde p}_1=p_1(a_1), \ {\tilde p}_2=p_1(a_2),\dotsc,{\tilde p}_{N}=p_1(a_N),\nonumber\\ \ {\tilde p}_{N+1}=p_2(a_1),\dotsc,{\tilde p}_{5N}=p_5(a_N). 
  \label{}
\end{align}
Let $C^{\rm data}_{\alpha \beta}$ be the corner of the covariance matrix corresponding to $\{{\tilde p}_\alpha \}$ that we have obtained earlier by inverting the total Fisher matrix. According to Bayes' theorem, the posterior probability distribution for the parameters $\{{\tilde p}_\alpha \}$ is a product of the likelihood and the prior probability. For Gaussian probabilities, this implies that the net covariance matrix (i.e. corresponding to the posterior probability) is the inverse of the sum of $\l[C^{\rm data}\r]^{-1}$ and $\l[C^{\rm prior}\r]^{-1}$. To construct the latter, one can follow the prescription in~\cite{Crittenden:2005wj,Crittenden:2011aa} and start with specifying a correlation function
\be
\langle (p_i(a) - p^{\rm lcdm}_i) (p_i(a') - p^{\rm lcdm}_i) \rangle \equiv \xi^{(i)}(|a-a'|)
\ee
where $p^{\rm lcdm}_i$ are the constant values of $p_i(a)$ in LCDM ($p_1=p_4=1$ and $p_2=p_3=p_5=0$), and where the form of the functions $\xi^{(i)}(|a-a'|)$ is chosen so that 
\ba
\nonumber
\xi^{(i)}(\Delta a) \rightarrow \xi^{(i)}(0)  \ &{\rm when}& \ \  \Delta a \equiv |a-a'| \ll a_c \\
\xi^{(i)}(\Delta a) \rightarrow 0 \ &{\rm when}& \ \  \Delta a \gg a_c  \ ,
\nonumber
\ea
where $a_c$ is the correlation length and $\xi^{(i)}(0)$ is a positive constant\footnote{In certain cases, it may be more appropriate to specify the correlation between points in $\log a$ rather than in $a$.}. One has to specify the functional form of $\xi^{(i)}(\Delta a)$ and we refer the reader to~\cite{Crittenden:2011aa} for an extended discussion of different choices. The choice does not make a big difference in practice. Using $\xi^{(i)}(\Delta a)$, one can calculate the prior covariance matrix for the binned $p_i(a_j)$ via
\be
C^{(i)}_{jk} = {1\over (\delta a)^2} \int_{a_j}^{a^j+\delta a} da \int_{a_k}^{a_k+\delta a} da' \xi^{(i)}(|a-a'|) \ ,
\ee
where $\delta a$ is the width of a bin in $a$. One can define such an $N\times N$ matrix for each of the five functions and combine them to form a block diagonal $5N \times 5N$ matrix $C^{\rm prior}$ for the parameters ${\tilde p}_\alpha$ such that
\be
C^{\rm prior}= {\rm diag}[C^{(1)},\dotsc,C^{(5)} ] \ .
\ee
This prior assumes that the five functions are independent of each other, which is true in the most general case but not in many specific models. For example, in $f(R)$ only one function is independent, while in more general Brans-Dicke models there are two. Such additional restrictions can be implemented, if desired, by adjusting the form of $C^{\rm prior}$.

Having constructed the prior matrix, one then compares the eigenmodes of $C^{\rm data}$ to the eigenmodes of the inverse of $\l[C^{\rm data}\r]^{-1}+\l[C^{\rm prior}\r]^{-1}$. The eigenmodes that are common to both, \ie those that survive, can be considered to be informative {\it with respect to that prior}. Due to the nature of the prior, one expects that slowly varying eigenmodes that are best constrained by data will have a higher chance to survive, while the high frequency modes will be suppressed. Naturally, the outcome of this comparison depends on the parameters of the prior, $a_c$ and $\xi^{(i)}(0)$. Their choice should, in principle, come from our theoretical prejudice. In practice, they can be tuned so that eigenmodes with variations on time scales comparable to Hubble time (or another time-scale that is theoretically motivated) survive. Thus, a PCA forecast is a key step for tuning the prior that can later be used in fitting to data, as we discuss next.

\subsection{Fitting to real data}

In a Fisher forecast, there is no limitation on making the number of bins $N$ as large as needed to achieve a convergence of well-constrained eigenmodes to the continuous limit. But this is not the case when fitting to real data, which involves searching for the maximum of a multi-parameter likelihood surface. The search stalls if the parameter space contains flat directions corresponding to nearly degenerate combinations of parameters, which is guaranteed to be the case when the number of bins is large. Fitting only the few best constrained eigenmodes amounts to a rather strong assumption that the amplitudes of the poorly constrained modes are known to be exactly zero, which amounts to adopting a strong, yet somewhat obscure, prior. Instead, in~\cite{Crittenden:2011aa,Zhao:2012aw} it was suggested to use the explicit smoothness prior described in the preceding subsection to aid the convergence of Monte Carlo Markov chains (MCMC).  This is achieved in practice by adding a term
\be
\chi^2_{\rm prior}=({\bf {\tilde p}}-{\bf {\tilde p}}^{\rm lcdm})^T [C^{\rm prior}]^{-1} ({\bf {\tilde p}}-{\bf {\tilde p}}^{\rm lcdm})
\ee
to the $\chi^2_{\rm data}$ in MCMC. The number of bins that one fits to data need not be very large. One typically needs a couple of bins per effective correlation length set by the prior. It remains to be shown for specific experiments how strong the prior needs to be in order for MCMC to converge. Once MCMC has converged, one can quantify the statistical significance of the detection of a departure from LCDM from the improvement in the $\chi^2_{\rm data}$. One can also compute the evidence for the best fit model and the Bayes' factors, since the prior probability is explicitly known. An explicit illustration of such a calculation for the case of $w(a)$ can be found in~\cite{Zhao:2012aw}.

We do not expect the reconstructed shapes of the individual functions $p_i(a)$ to be highly informative because observables will only constrain their combinations and degeneracies between parameters $\{{\tilde p}_\alpha \}$ will make marginalized errors on them large. One could instead use Eqs.~(\ref{eq:gamma}) and (\ref{eq:mu}) to visualize reconstructions of $\mu$ and $\gamma$, or $\mu$ and $\Sigma$, as surfaces in the $(a,k)$ space.

\section{Summary}
\label{sec:summary}

In this paper we have motivated a parametric form for the modified growth functions $\mu$ and $\gamma$ that fixes their scale dependence to a ratio of polynomials in $k$ and have shown that generally the denominator of $\gamma$ is equal to the numerator of $\mu$. We arrive at this form by taking the quasi-static approximation (QSA) in the equations for scalar perturbations derived from a covariant action that allows for modifications of gravity and any number of scalar degrees of freedom. We examine the impact of assuming the QSA in our derivation and conclude that, until a viable counterexample is found, the non-quasi-static effects of modified gravity can be assumed to be negligible. We nevertheless note that the final form of our parametrization allows for a detection of some non-quasi-static signatures, but not necessarily for the most general ones in that regime. 

We further argue that for most of the viable modifications of gravity discussed in the literature, the polynomials in $k$ are even and of second degree, effectively reducing the number of time-dependent coefficients to five. Since these coefficients are functions of the background variables only, they can be safely assumed to be slowly varying. This justifies using an explicit smoothness prior on their shape when fitting them to data, similarly to the Bayesian framework developed for the reconstruction of $w(a)$ in~\cite{Crittenden:2011aa,Zhao:2012aw} .

A forecast of future reconstructions of $\mu$ and $\gamma$ based on this approach is currently in progress~\cite{in-progress}.

\acknowledgments We benefited from discussions and previous collaborations with Rob Crittenden, Kazuya Koyama and Gong-Bo Zhao, as well as useful interactions with Tessa Baker, Antonio de Felice, Pedro Ferreira, and Constantinos Skordis. AS is supported by the SISSA Excellence Grant and acknowledges the use of DAMTP/SISSA collaboration grant. LP is supported by an NSERC Discovery grant.

\end{document}